\documentclass[3p]{elsarticle}

\usepackage{amsfonts}
\usepackage{amsmath, enumerate}
\allowdisplaybreaks[4]
\usepackage{epstopdf}
\usepackage{hyperref}
\usepackage{enumerate}
\usepackage{graphicx}
\usepackage{amssymb,epsfig,times,bbm,lineno,color}

\usepackage{exscale}
\usepackage{relsize}
\usepackage{subfigure}
\usepackage{colortbl,dcolumn}
\usepackage{amsthm}
\usepackage{cases}%

\newtheorem{theorem}{Theorem}

\theoremstyle{remark}
\newtheorem{remark}{\emph{\textbf{Remark}}}
\newtheorem*{notation}{\emph{\textbf{Notations}}}
\newtheorem{definition}{\emph{\textbf{Definition}}}
\newtheorem{lemma}{\emph{\textbf{Lemma}}}

\newtheorem{exa}{\emph{\textbf{Example}}}

\begin{document}\pagestyle{empty}
\begin{frontmatter}

\title{Feedback  pinning control of collective behaviors aroused by epidemic spread on complex networks}

\author[SHU]{Pan Yang}
\author[SHU]{Zhongpu Xu}
\author[SZU]{Jianwen Feng}
\author[SHU]{Xinchu Fu\corref{cor1}}
\ead{xcfu@shu.edu.cn}
\address[SHU]{Department of Mathematics, Shanghai University, Shanghai 200444, China}
\cortext[cor1]{Corresponding author}
\address[SZU]{College of Mathematics and Statistics, Shenzhen University, Shenzhen 518060, China}

\begin{abstract}
This paper investigates epidemic control behavioral synchronization for a class of complex networks
resulting from spread of epidemic diseases via pinning feedback control strategy. Based on the quenched
mean field theory, epidemic control synchronization models with inhibition of contact behavior is
constructed, combining with the epidemic transmission system and the complex dynamical network carrying
extra controllers. By the properties of convex functions and \emph{Gerschgorin} theorem, the epidemic
threshold of the model is obtained, and the global stability of disease-free equilibrium is analyzed. For
individual's infected situation, when epidemic spreads, two types of feedback control strategies
depended on the diseases' information are designed: the one only adds controllers to infected
individuals, the other adds controllers both to infected and susceptible ones. And by
using \emph{Lyapunov} stability theory, under designed controllers, some criteria that guarantee
epidemic control synchronization system achieving behavior synchronization are also derived.
Several numerical simulations are performed to show the effectiveness of our theoretical results.
As far as we know, this is the first work to address the controlling behavioral synchronization
induced by epidemic spreading under the pinning feedback mechanism. It is hopeful that we may have more deeper
insight into the essence between disease's spreading and collective behavior controlling in complex
dynamical networks.
\end{abstract}


\end{frontmatter}

\section{Introduction}

Complex networks are ubiquitous around the world, such as the World Wide Web, biological
networks, food chain networks, power grids and transportation networks, economic and social
networks, see~\cite{paper1}-\cite{paper10}
and references therein. Since complex networks
can be used as models to reflect certain intrinsic natures in the real world, a lot of
achievements have been made on complex networks~\cite{paper11,paper12}.
Synchronization and propagation, two typical interesting and important dynamical behaviors
in networks, have attracted much attention from various fields in recent years, including
communication, engineering, physical science, mathematics, sociology and biology.
On the one hand, many kinds of synchronization
mechanisms have been discovered over the last decades, such as the complete
synchronization~\cite{paper13,paper14}, the cluster
synchronization~\cite{paper15,paper16}, the lag
synchronization~\cite{paper17,paper18},
and so on, see~\cite{paper19}-\cite{paper21} and references therein. On the other hand,
the spreading of epidemic disease in complex networks have also been received more
and more attentions due to the development of network science research since 1998.
When an epidemic disease spreads in a realistic network, the information about
the disease transfer only along the edges, which means that there will no direct transmission
in any pairs of individuals without relationship between them.
Taking into consideration of the significance about spreading of epidemic
diseases on complex networks, many successful works have been devoted by a lot of
researchers from mathematical biology field~\cite{paper22}-\cite{paper24}. The
problem about global stability and attractivity of a network-based SIS epidemic
model with nonmonotone incidence rate was investigated in~\cite{paper25}, and the
conditions guaranteeing global asymptotical stability or global
attractivity of an endemic equilibrium are also obtained.
And the issue of network-level reproduction number and
extinction threshold for vector-borne diseases was studied in~\cite{paper26},
meanwhile relationships between the basic reproduction numbers and the extinction
thresholds of corresponding models are obtained by using some assumptions.

One may think that collective behaviors and spreading behaviors are two
independent patterns of behavior in complex networks, however, the fact seem to
be not the case. In fact, when an epidemic disease spreads in a network,
individuals always change their behaviors to protect themselves from being infected
by infected individuals around them. They may take steps consistently, such as washing their hands
with soaps frequently, having the cure in hospital, avoiding going to the crowded and
doing exercise to strengthen their resistance to infection, and so on,
to combat the disease. These consistent
behaviors against the epidemic disease are also called as behavior synchronization
induced by contagious diseases. There are some successful works
in recent years~\cite{paper27}-\cite{paper30} taking the
synchronization and epidemic transmission in complex networks
into consideration simultaneously.
Motivated by these interesting works, in this paper, we will further study the problem about the behavior synchronization induced by epidemic spread.

In~\cite{paper31}, authors considered adaptive mechanism between dynamical
synchronization and epidemic spreading behavior on complex networks, in which two models of
epidemic synchronization were introduced for the first time to analyze the stability of epidemic
synchronization and obtained local and global conditions. Following that, the interplay
between collective behavior and spreading dynamics on complex networks was further investigated
in~\cite{paper32}, where several bidirectional networks models of spreading phenomena
were constructed, and found that the collective behavior and spreading behavior
are influenced each other. But the models in the above two papers are all based on the
heterogeneous mean-field theory. In order to be more accurate in predicting epidemic
synchronization, authors in~\cite{paper33} introduced the quenched mean-field theory to
construct three epidemic synchronization models without delays or with distinct delays,
and got some local and global conditions to guarantee the results. Therefore, the quenched
mean-field theory will be also considered in our work for more precise prediction of epidemic
collective behaviors. However, models in these works are all about adaptive synchronization
with adaptive exterior coupling strength. As is known to all, it is relatively difficult
to synchronize a network only depending on itself evolution. Hence, we will
add external controllers to achieve behavioral synchronization aroused by the epidemic
disease transmission in this work.

It is acknowledged that adding external controllers to reach synchronization
is an important and effective method in a scale-large complex network~\cite{paper34}-\cite{paper36}.
When an epidemic disease breaks out, it is also important to actively control collective
behaviors, as the individuals may not get full information about the diseases or even do
not know the event of contagion in the early stage. Therefore, in order to make individuals
in a network response
quickly in the early stage of an epidemic disease spreading, some external controlling methods are
carried out to reach behavioral consistency for the individuals, named as controlling
behavioral synchronization induced by epidemic disease transmission. Such control can be performed via
broadcasting and other media of information dissemination or some certain well-directed means,
by which individuals may consistently behavior, such as going to the hospital, washing their hands with
soaps, etc., to protect themselves from being infected, and resist the diffusion
of contagions. As far as we know, there are no works concerning epidemic
behavior synchronization via a control strategy. Therefore, we will discuss the collective
behaviors induced by epidemic disease transmission in the view of external pinning
feedback controller introduced in~\cite{paper34} with the normalized outer-coupling strength
in the synchronous models. The difference in this case is that the pinning feedback
controllers depend upon the information about epidemic spreading, where controllers work
only when contagious disease spreads. In addition, noticing a realistic fact that
there are always the inhibition of contact behaviors among individuals rooted in the
subconscious, especially when an contagion spreads. In other words, individuals may change
their willing to contact with others relying on the risk of being infected, and the consistency
with the behavioral target resulting from protecting themselves, which can be reflected as
the higher of the risk of being infected, the less the individuals are willing to contact
with each other. And
the less distinct between behavioral targets, persons are desired to touch with others.
Thus it is natural to consider the inhibition of contact behavior into our models.

Inspired by the above discussions, this paper further studies the the pinning feedback
control of behavior synchronization induced by epidemic diseases spreading. The
outer-coupling strength is normalization and the external pinning feedback controllers are
equipped in the synchronous models. And in the epidemic spreading models, the quenched
mean-field theory is introduced for more accurate reflection of the realistic situations, and
more precise evaluation of the control behavior synchronization aroused from transmission of
contagious. What's more, in order to simulate more real human reaction, the inhibition of
contact behaviors are also considered in our epidemic spreading models. That is to say,
on the one hand, the higher the risk of being infected, the less the willing to contact with each other.
On the other hand, the more consistent the individuals' behaviors are, the stronger the desires to keep in touch. We will use distinct pinning feedback control strategies to achieve
corresponding control of behavior synchronization, which means the corresponding controllers
will be added to only a portion of nodes or individuals for each strategy in the network.

The main contributions of this work are listed as follows. Firstly, combined with the
synchronous systems and Susceptible-Infected-Susceptible (SIS) epidemic systems, based on
the quenched mean-field theory, we construct a class of models of epidemic controlled behavioral
synchronization with the external controllers, in which the outer-coupling strength is
normalized in the synchronous system. Secondly, in order to perform more real simulation, the
inhibition of contact behavior is concerned in the epidemic system, which means that
the desires to contact with others not only rely on the risk of being infected, but also
depend on the difference between behavioral target states. Finally,
pinning feedback control are equipped into the epidemic synchronization for the first time.
In other words, only a part of nodes will be controlled by corresponding feedback
strategies, and the controllers are related to the information of the epidemic spread,
which means the controllers will work only when the epidemic spreads, or failures
with the epidemic extinction.

The rest of the paper is organized as follows. In Section~\ref{mathemodel}, the problem
is formulated and some helpful definitions, lemmas, assumptions and notations are given.
In Section~\ref{mainresults}, the existence of endemic equilibrium and the global
stability of disease-free equilibrium are analyzed at first in the epidemic
controlling synchronization models, and then the theoretical results for distinct controlled
behaviors synchronization induced by epidemic spreading under different control strategies
are derived. Some numerical examples are shown to illustrate the theoretical results
in Section~\ref{simulation}. Finally, the conclusions of this paper are presented in
Section~\ref{conclusions}.

\begin{notation}
Throughout the paper, $\mathbb{R}$ is the set of real numbers, $\mathbb{R}^{n}$ represents
the $n$ dimensional Euclidean space, $\mathbb{R}_{+}^{n}$ is the $n$ dimensional
vector space with positive elements, and $\mathbb{R}^{n\times m}$ denotes the set of real
matrices with order~$n\times m$. The transposition for a vector or a matrix is
represented by superscript $T$. $\lambda_{\min}(\cdot)$ and $\lambda_{\max}(\cdot)$
represent the minimum and maximum eigenvalues of a square matrix, respectively. And $I_{N}$
is the identity matrix of order~$N$. $\|\cdot\|$ represents the Euclidean norm for vectors
and the induced $2$-norm for matrices. And $\otimes$ is the Kronecker product.
For a matrix $P$, $P>0 (<0)$ means that $P$ is positive (negative) definite.
\end{notation}

\section{Models and Preliminaries}\label{mathemodel}

The mathematical models about infectious disease dynamics and complex dynamical network systems are
given in this part. And some fundamental conceptions and necessary preliminaries are introduced
for later use.

\subsection{Mathematical models}

\subsection*{{A. Complex dynamical network model}}

We now introduce a coupled complex network with $N$ nodes or individuals
under controlling
\begin{align}\label{synmodel}
\dot{x}_{i}(t)=f(x_{i}(t))+\sum_{j=1}^{N}a_{ij}x_{j}(t)+u_{i}(t),\quad i=1,\cdots,N,
\end{align}
where $x_{i}(t)=(x_{i1}(t),\cdots,x_{in}(t))^{T}\in{\mathbb{R}^{n}}$ represents the $i$-th
node's state, $f(\cdot):$ $\mathbb{R}^{n}\rightarrow\mathbb{R}^{n}$
is a differentiable function that represents the behavior of individual system,
and $f(x_{i}(t))=(f_{1}(x_{i}(t)),f_{2}(x_{i}(t)),\cdots,f_{n}(x_{i}(t)))^{T}\in\mathbb{R}^{n}.$  Matrix $A=(a_{ij})\in{\mathbb{R}^{N\times N}}$ represents the out-coupled structure,
in which $a_{ij}=a_{ji}=1$ if there is a connection between nodes (or individuals) $i$
and $j$, and $a_{ij}=a_{ji}=0$ otherwise. $u_{i}(t)$ is the external controller for each
node $i$, which will be defined later.

Define the $\emph{Laplacian}$ matrix as $L=(l_{ij})\in{\mathbb{R}^{N\times N}}$:
$$l_{ij}=-a_{ij},$$ $$l_{ii}=-\sum_{j=1,j\neq i}^{N}l_{ij},\quad i=1,2,\cdots,N$$.

Assume the $\emph{Laplacian}$ matrix $L$ is irreducible, which indicates that network
with $N$ nodes is strongly connected, e.g., there exists one path from any node $i$ to $j.$
From the definition of $\emph{Laplacian}$ $L$, it is obvious that $L$ has one zero
eigenvalue and $N-1$ positive eigenvalue, which means that there exists a unitary matrix $U$
such that
$$ U^{T}LU={\rm diag}\{\lambda_{1},\lambda_{2},\cdots,\lambda_{N}\},$$
we sort these eigenvalues as $0=\lambda_{1}\leq\lambda_{2}\leq\cdots\leq\lambda_{N}.$

\subsection*{{B. SIS epidemic dynamics model}}

Taking into account the network's topological structure, this paper introduces the following
quenched networks for infectious disease systems (or quenched mean field model):
\begin{subequations}\label{infemodel}
\begin{numcases}{}
\dot{\rho}_{i}(t)=-\rho_{i}(t)+\lambda\varphi_{i}(t)[1-\rho_{i}(t)]\sum_{j=1}^{N}a_{ij}\rho_{j}(t)\label{infemodel:1}\\
\dot{S}_{i}(t)=\rho_{i}(t)-\lambda\varphi_{i}(t)[1-\rho_{i}(t)]\sum_{j=1}^{N}a_{ij}\rho_{j}(t)\label{infemodel:2}
\end{numcases}
\end{subequations}
where $\rho_{i}(t)$ and $S_{i}(t)$ represent the infected probability and the uninfected
probability of node $i$ at time $t$, respectively. $\lambda$ is the infection rate.
$\varphi_{i}(t)$ is the probability of inhibition to contact behavior for each individual $i$,
which has the representation as $\varphi_{i}(t)=1-\gamma_{i}(t)$, where $\gamma_{i}(t)$ is defined
as follows,
$$\gamma_{i}(t)=\alpha\rho_{i}(t)\frac{e_{i}^{T}(t)e_{i}(t)}{\sum_{i=1}^{N}e_{i}^{T}(t)e_{i}(t)},$$
in which $\alpha$ denotes the gain of desire for contact, and $\alpha\in(0,1).$ And the $e_{i}(t)$
is the synchronization errors  for the $i$-th node defined later.

\begin{remark}
Note that the inhibition to contact behavior is a natural reaction for all the
population, especially in the case of breaking out of an epidemic disease.
This behavior not only depends on
the information about the risk of being infected by other
individuals, but also on the target to reach.
On the one hand, the greater the risk of being infected is,
the weaker the desire to contact with others is.
On the other hand, the smaller the behavior error between synchronous target is,
the stronger the willing to contact with others is.
\end{remark}

\subsection*{{C. SIS epidemic controlling synchronization model}}

Then by combining the SIS epidemic dynamics model and complex dynamical network model,
we get the epidemic controlling synchronization model with external controllers:
\begin{subequations}\label{infenetmodel}
\begin{numcases}{}
\dot{x}_{i}(t)=f(x_{i}(t))-\sum_{j=1}^{N}l_{ij}x_{j}(t)+u_{i}(t)\label{infenetmodel:1}\\
\dot{\rho}_{i}(t)=-\rho_{i}(t)+\lambda\varphi_{i}(t)[1-\rho_{i}(t)]\sum_{j=1}^{N}a_{ij}\rho_{j}(t)\label{infenetmodel:2}
\end{numcases}
\end{subequations}

This kind of compound model mixes dynamical equation of individual's state and
equation for individual's spreading of disease, and in the case of the controlling
behavioral synchronization induced by the transmission of epidemic disease, $x_{i}(t)$
denotes the state variable of the individual (or node) $i$ at time $t$ in the epidemic
disease network, which reflects certain daily behavioral patterns (washing the hands,
traveling and having a rest, etc.). $\rho_{i}(t)$ represents the probability of being
infected for individual $i$ at time $t$. Assume that individuals may be infected only
when there exists connections among them. At this moment, model~\eqref{synmodel} and
model~\eqref{infemodel} share the same topology structure. In other words, the dynamical
behaviors of individuals and the epidemic spreading behaviors appear simultaneously in the same
network, which indicates that the process of individuals' dynamical behaviors will
accompany the spreading behavior of epidemic disease in the disease network. And the $f$
defines the local dynamics that represents the summation of distinct behaviors
for epidemic individuals, and the $f$ is active, dynamic as well as complex. In this paper,
we realize the coupling between the model~\eqref{synmodel} and model~\eqref{infemodel} by
external controllers depending on the epidemic information of individuals' disease and
the synchronous errors, then reaching the point of controlling behavioral synchronization
under the premise of epidemic diseases transmission.

In the first equation of model~\eqref{infenetmodel}, $u_{i}(t)$ represents the external
controller for each node $i$ consisting of two parts, $\rho_{i}(t)$ and $e_{i}(t)$, in
which $\rho_{i}(t)$ can be regarded as the information about the epidemic disease for
each individual, while the other term is the essential part in the pinning control
strategy that measures the synchronous state of each individual (or node). Then the
controllers will be activated and play a role according to the situation about the
breaking-out of the epidemic disease under this framework for controllers. Which means that
the controllers will work to lead the behavioral synchronization when the epidemic
disease spreads, e.g., $\rho_{i}(t)$ will not be zero as $t$ tends to
positive infinity. When the epidemic disease dose not spread, the controllers
will be invalid due to the disease's die out, i.e., $\rho_{i}(t)$ goes to zero.
Hence it is no needed to control individuals in the network to get the collective
behavior. Thus it is meaningful to design the pinning
controllers to relate to $\rho_{i}(t)$.

\subsection{Preliminaries}

For analytical requirement, some useful preliminaries are introduced in this part.

\begin{definition}\label{synchdef}
A complex dynamical network system containing $N$ individuals (or nodes) is said to reach
asymptotical synchronization, if for each pair nodes $i$ and $j$ ($i\neq j$)
we have
$$
\lim_{t\rightarrow \infty}\|x_{i}(t)-x_{j}(t)\|=0.
$$
Assume that there exists a synchronous orbit $s(t),$  then it is just needed to check $\|x_{i}(t)-s(t)\|\rightarrow 0$ as $t\rightarrow\infty$ to achieve asymptotical synchronization.
As for the case of distinct synchronous orbits $s_{k}(t),$ if $\|x_{i}(t)-s_{k}(t)\| \rightarrow 0$
 as $t\rightarrow\infty,$ $i$ belongs to distinct groups $k$, then the dynamical network achieves cluster synchronization.
\end{definition}

\begin{definition}[\cite{paper38}]\label{QUAD}
A continuous function $f(t,x):[0,+\infty]\times \mathbb{R}^{n}\rightarrow\mathbb{R}^{n}$
is said to be in a \rm{QUAD} function class, denoted as $f\in \mbox{QUAD}(P,\Delta,\eta)$, if there exists
a positive definite diagonal matrix $P=diag\{p_{1},p_{2},\cdots,p_{n}\},$ a diagonal matrix $\Delta=diag\{\delta_{1},\delta_{2},\cdots,\delta_{n}\}$ and a constant $\eta>0,$ such that $f$
satisfies following condition:
$$(x-y)^{T}P(f(t,x)-f(t,y)-\Delta(x-y))\leq -\eta(x-y)^{T}(x-y)$$
for all $x,y\in\mathbb{R}^{n}.$
\end{definition}

\begin{remark}
Notice a fact that there are many chaotic systems having the properties of the {QUAD}
function class, for example, the Lorenz system, the R\"{o}ssler system, the Chen system,
the Chua's circuit, and the logistic differential system.
\end{remark}

\begin{definition}
Matrix $L$ is said to be class $A1,$ denoted as $L\in A1,$ if the following conditions
hold:
\begin{enumerate}
  \item $l_{ij}\leq0,i\neq j,l_{ii}=-\sum_{j=1,j\neq i}^{N}l_{ij},$
  \item $L$ is irreducible.
\end{enumerate}
\end{definition}

If $L$ is symmetric, we say $L$ belongs to class $A2,$ denoted as $L\in A2.$ We can easily
obtain that if $L$ is class $A1,$ then $L$ has the property of zero row-sum,
e.g., $\sum_{j=1}^{N}l_{ij}=0.$

As for a rectangular matrix $L\in \mathbb{R}^{N_{1}\times N_{2}},$ if its each row-sum is
zero, e.g., $\sum_{j=1}^{N_{2}}l_{ij}=0,i=1,2,\cdots,N_{1},$ then we call $L$ is in
class $A3,$ denoted as $L\in A3.$

\begin{lemma}[\cite{paper39}]\label{L1}
For any vectors $x,y\in\mathbb{R}^{n},$ there exists positive $\theta,$ such that
$$2x^{T}y\leq \frac{1}{\theta}x^{T}x+\theta y^{T}y.$$
\end{lemma}

\begin{lemma}[\emph{Gerschgorin} Theorem~\cite{paper40}]
Let $A=(a_{ij})\in\mathbb{C}^{n\times n}$ and let $r_{i}=\sum_{ j=1, j\neq i}^{n}|a_{ij}|, ~i=1,2,\cdots,n.$
Then all the eigenvalues of $A$ lie in the union of $n$ closed discs $\bigcup_{i=1}^{n}\{z\in\mathbb{C}:|z-a_{ii}|\leq r_{i}\},$ where $\mathbb{C}$ is the
set of complex numbers and $\mathbb{C}^{n\times n}$ represents the complex matrices set
of order $n\times n.$
\end{lemma}


\section{Main Results}\label{mainresults}

The main concern of our study is controlling the
behavioral dynamics system~~\eqref{infenetmodel} to
achieve desired behavior synchronization via designed
controller $u_{i}(t)$ during the spread of an infectious disease.

The reason to add controllers can be explained as follows.
Because of unresponsive information transmission or unawareness about the disease type,
when the infectious disease outbreaks, e.g., infection rate is larger than
the epidemic threshold, people may not make corresponding response to this case in a short
time, or the information about disease is transmitted by medical institutions to the public
after some research. At this moment, some guidance or measures should be carried out on their
behaviors to reach an identical action, including getting cure in hospital and isolating,
or washing hands frequently and having rest and so on, which prevent a disease from next
outbreak. Obviously, this kind of behavioral consistency is resulted from the disease
transmission, and is achieved under external feedback controllers.

Before analyzing behavior synchronization, the model~\eqref{infemodel}'s equilibrium will
be studied first.

\subsection{Analysis of the infectious disease system}

Let $$\mathbf{\Gamma}={\{(\rho_{i},\cdots,\rho_{N})\in\mathbb{R}_{+}^{N}|0\leq \rho_{i}\leq 1,~i=1,\cdots,N\}}.$$
Obviously, it is easy to know that $\Gamma$ is the positive invariant set for
system~\eqref{infemodel}.

\begin{theorem}\label{pequil}
If the infection rate  $\lambda>\frac{1}{(1-\alpha)\rho(A)}$, then the epidemic dynamics system~\eqref{infemodel} has a positive equilibrium $E^{*}=(S_{1}^{*},\rho_{1}^{*},\cdots,S_{N}^{*},\rho_{N}^{*}),$ i.e., the endemic
equilibrium.

\begin{proof}
Denote by $\rho_{i}$ the steady state of $\rho_{i}(t)$, and let the right-hand side equal to zero
in the first one equation~\eqref{infemodel:1} of the system~\eqref{infemodel}, then
define
\begin{align}
\Theta=\sum_{j=1}^{N}a_{ij}\rho_{j}.
\end{align}
From the system~\eqref{infemodel}, we have
\begin{align}
\Theta=\sum_{j=1}^{N}a_{ij}\frac{\lambda\varphi_{i}\Theta}{1+\lambda\varphi_{i}\Theta}.
\end{align}
In order to analyze the $\Theta$'s property, construct the auxiliary function
\begin{align}
F(\Theta)=\Theta-\sum_{j=1}^{N}a_{ij}\frac{\lambda\varphi_{i}\Theta}{1+\lambda\varphi_{i}\Theta}.
\end{align}
Then, we can get the first-order derivative at $\Theta$
\begin{align}
\frac{dF(\Theta)}{d\Theta}=1-\sum_{j=1}^{N}a_{ij}\frac{\lambda\varphi_{i}}{(1+\lambda\varphi_{i}\Theta)^{2}},
\end{align}
and the second-order derivative at $\Theta$
\begin{align}
\frac{d^{2}F(\Theta)}{d\Theta^{2}}=\sum_{j=1}^{N}a_{ij}\frac{2\lambda^{2}\varphi_{i}^{2}}{(1+\lambda\varphi_{i}\Theta)^{3}}>0,
\end{align}
As for the first-order derivative at $\Theta=0$, we have
\begin{align*}
\frac{dF(\Theta)}{d\Theta}\Big|_{\Theta=0}&=1-\sum_{j=1}^{N}a_{ij}\lambda\varphi_{i}
<1-\sum_{j=1}^{N}a_{ij}\lambda(1-\alpha).
\end{align*}
Then, it is necessary for the equation $F(\Theta)=0$
to have a unique positive solution within the internal $(0,1)$ that the following
relation holds
\begin{align}\label{lemlarge}
\lambda>\frac{1}{(1-\alpha)\sum_{j=1}^{N}a_{ij}},
\end{align}
and inequality~\eqref{lemlarge} means
\begin{align}
\lambda>\frac{1}{(1-\alpha)\max_{i}\sum_{j=1}^{N}a_{ij}}.
\end{align}

On the other hand, according to the \emph{Gersgorin} disc theorem,
\begin{align}
|\lambda_{i}(A)|\leq \sum_{j=1}^{N}a_{ij},
\end{align}
so we further know that
\begin{align}
\max_{i}|\lambda_{i}(A)|\leq \max_{i}\sum_{j=1}^{N}a_{ij}.
\end{align}
From the definition of spectral radius for the matrix $A$, we have
\begin{align}
\rho(A)=\max_{i}|\lambda_{i}(A)|\leq \max_{i}\sum_{j=1}^{N}a_{ij}.
\end{align}
Therefore, by the condition
\begin{align}
\lambda>\frac{1}{(1-\alpha)\rho(A)},
\end{align}

we obtain the first-order derivative of $F(\Theta)$
\begin{align*}
\frac{dF(\Theta)}{d\Theta}\Big|_{\Theta=0}<0.
\end{align*}
Then the proof is completed.
\end{proof}
\end{theorem}

It is worthy to notice that Theorem~\ref{pequil} discovers when the infection rate
  $\lambda>\frac{1}{(1-\alpha)\rho(A)},$ there
exists $\rho_{i}^{*}\in(0,1),i=1,2,\cdots,N$, such that
\begin{align*}
\lim_{t\rightarrow \infty }\rho_{i}(t)=\rho_{i}^{*},\quad i=1,2,\cdots,N.
\end{align*}

Next, the global stability of disease-free equilibrium  will be considered. Some
mathematical expressions are given for analytical requirement.
Setting the positive left eigenvector corresponding to the maximum eigenvalue
of $A$ as $(\omega_{1},\cdots,\omega_{N})^{T}\in\mathbb{R}^{N},$ which indicates
that
$$(\omega_{1},\cdots,\omega_{N})A=(\omega_{1},\cdots,\omega_{N})\rho(A).$$

\begin{theorem}\label{freestable}
The disease-free equilibrium in the epidemic dynamics system~\eqref{infemodel}
is globally asymptotically stable if the infection rate
 $\lambda\leq\frac{1}{(1+\alpha)\rho{(A)}}.$
While it is unstable when $\lambda>\frac{1}{(1-\alpha)\rho{(A)}}.$

\begin{proof}
Choose the following \emph{Lyapunov} functional
\begin{align}
V(t)=\sum_{i=1}^{N}\omega_{i}\rho_{i}(t).
\end{align}
Differentiating the function $V(t)$ along the trajectories of Eq.~\eqref{infemodel}, we obtain
\begin{align*}
\dot{V}(t)&=\sum_{i=1}^{N}\omega_{i}\dot{\rho}_{i}(t)\\
&=\sum_{i=1}^{N}\omega_{i}\Big[-\rho_{i}(t)+\lambda\varphi_{i}(t)(1-\rho_{i}(t))\sum_{j=1}^{N}a_{ij}\rho_{j}(t)\Big]\\
&=-\sum_{i=1}^{N}\omega_{i}\rho_{i}(t)+\lambda\sum_{i=1}^{N}\sum_{j=1}^{N}\omega_{i}\varphi_{i}(t)a_{ij}\rho_{j}(t)
-\lambda\sum_{i=1}^{N}\sum_{j=1}^{N}\omega_{i}\rho_{i}(t)\Big[1-\alpha\rho_{i}(t)\frac{e_{i}^{T}(t)e_{i}(t)}{\sum_{i=1}^{N}e_{i}^{T}(t)e_{i}(t)}\Big]a_{ij}\rho_{j}(t)\\
&\leq\Big[\lambda(1+\alpha)\rho(A)-1\Big]\omega\rho,
\end{align*}
where $\omega=(\omega_{1},\cdots,\omega_{N})^{T}\in\mathbb{R}^{N},$ $\rho=(\rho_{1}(t),\cdots,\rho_{N}(t))^{T}\in\mathbb{R}^{N}.$

Therefore, if $\lambda\leq\frac{1}{(1+\alpha)\rho{(A)}}$, we have $\dot{V}(t)\leq0$, and $\dot{V}(t)=0$ if and only if $\rho_{i}(t)=0, i=1,2,\cdots,N.$
By \emph{LaSalle's} Invariance Principle, disease-free equilibrium is globally asymptotically stable.

On the other hand, inequality relation $(1-\alpha)\rho(A)<(1+\alpha)\rho(A)$ indicates that $\frac{1}{(1-\alpha)\rho(A)}>\frac{1}{(1+\alpha)\rho(A)}.$ And from the Theorem~\ref{pequil}, which means that the $\dot{V}(t)>0$ while $\lambda>\frac{1}{(1-\alpha)\rho(A)}.$ That is to say, the disease-free equilibrium is unstable, and there is one positive equilibrium in the interior $\mathring{\mathbf{\Gamma}}$ of $\mathbf{\Gamma}.$
\end{proof}
\end{theorem}

\subsection{Synchronization on infectious disease complex network system with controllers }

To control the transmission of an infectious disease, we want to make the behavioral dynamics system~\eqref{infenetmodel} reach a suitable behavior synchronization under
the designed external controllers. We will achieve distinct synchronous types according
to selecting different control strategies on the epidemic controlling synchronization system.
Assuming that there exist synchronous orbits $s(t)$ and $s_{k}(t)$ for two distinct
synchronous types as the complete synchronization machine and the cluster
synchronization for two groups based on the information of individuals' disease,
respectively.

\subsection*{A. Control the infected individuals or nodes}

A well known fact indicates that control all nodes on complex network system is difficult
to implement and unrealistic, while it is effective and enforceable to control a portion
of nodes. Therefore, we design the following external controllers
\begin{align}\label{controller1}
u_{i}(t)=-d_{i}\rho_{i}(t)(x_{i}(t)-s(t))
\end{align}
in which $d_{i}$ is the control strength, and $d_{i}>0$ when $i=1,2,\cdots,r_{l}$, and $d_{i}=0$
otherwise.

\begin{remark}
Note that the designed controllers are only added to infectious individuals, under pinning
feedback controllers, the whole network containing infectious and susceptible individuals
achieves desired behavior synchronization.
\end{remark}

Let
$s(t)$ be the synchronous target that satisfies the uncoupled system as follows
\begin{align}
\dot{s}(t)=f(s(t)).
\end{align}

Define the errors between node and target as $e_{i}(t)=x_{i}(t)-s(t), i=1,2,\cdots,N.$
Then we can get the error system with controllers
\begin{align}\label{errors}
\dot{e}_{i}(t)=f(x_{i}(t))-f(s(t))-\sum_{j=1}^{N}l_{ij}e_{j}(t)-d_{i}\rho_{i}(t)e_{i}(t).
\end{align}

%

\begin{theorem}\label{synctheorem}
Assume the local dynamics $f$ belongs to the QUAD function class. When an infectious disease
spreads, which means $\lambda>\frac{1}{(1-\alpha)\rho(A)},$ then the endemic
equilibrium is globally  asymptotically stable. Under this condition,
the behavioral dynamics system~\eqref{infenetmodel} can achieve behavior
synchronization under the extra controllers~\eqref{controller1} if there exists a
positive $\eta>0$ and the following condition holds:
\begin{align}\label{condition1}
(-\eta-\lambda_{2}\lambda_{\min}(P))I_{Nn}+I_{N}\otimes(P\Delta)-D\otimes P<0,
\end{align}
where
\begin{align*}
D=diag\{(d_{1}(\rho_{1}^{*}-\varepsilon),\cdots,d_{r_{l}}(\rho_{r_{l}}^{*}-\varepsilon),0,\cdots,0)\}.
\end{align*}
\begin{proof}
Choose the functional
\begin{align}
V(t)=V_{1}(t)+V_{2}(t),
\end{align}
in which
\begin{align}\label{V1}
V_{1}(t)=\frac{1}{2}\sum_{i=1}^{N}e_{i}^{T}(t)Pe_{i}(t),
\end{align}
\begin{align}\label{V2}
V_{2}(t)=\sum_{i=1}^{N}c_{i}V_{2i},c_{i}\geq0,
\end{align}
where
\begin{align}
V_{2i}=S_{i}(t)-S_{i}^{*}-S_{i}^{*}\ln\frac{S_{i}(t)}{S_{i}^{*}}+\rho_{i}(t)-\rho_{i}^{*}-\rho_{i}^{*}\ln\frac{\rho_{i}(t)}{\rho_{i}^{*}}.
\end{align}

Obviously, $V_{1}(t)$ is a \emph{Lyapunov} function, while $V_{2}(t)$ needs to be
clarified further.
Next, differentiating the function $V(t)$ along the trajectories of Eq.~\eqref{infenetmodel}.

Firstly, according to the error systems~\eqref{errors}, we can get the differential form
\begin{align*}
\dot{V}_{1}(t)&=\sum_{i=1}^{N}e_{i}^{T}(t)P\dot{e}_{i}(t)\\
&=\sum_{i=1}^{N}e_{i}^{T}(t)P\Big[f(x_{i}(t))-f(s(t))-\sum_{j=1}^{N}l_{ij}e_{j}(t)-d_{i}\rho_{i}(t)e_{i}(t) \Big]\\
&\leq-\eta\sum_{i=1}^{N}e_{i}^{T}(t)e_{i}(t)+\sum_{i=1}^{N}e_{i}^{T}(t)P\Delta e_{i}(t)-\sum_{i=1}^{N}\sum_{j=1}^{N}e_{i}^{T}(t)Pl_{ij}e_{j}(t)-\sum_{i=1}^{N}e_{i}^{T}(t)Pd_{i}\rho_{i}(t)e_{i}(t).
\end{align*}

Let $y(t)=(y_{1}^{T}(t),y_{2}^{T}(t),\cdots,y_{N}^{T}(t))^{T}=(U^{T}\otimes I_{N})e(t),$ where $U$ is the unitary matrix that satisfies $U^{T}U=UU^{T}=I$. Then, we have
\begin{align*}
\sum_{i=1}^{T}\sum_{j=1}^{N}e_{i}^{T}(t)Pl_{ij}e_{j}(t)&=e^{T}(t)(L\otimes P)e(t)\\
&=y^{T}(t)(\Lambda\otimes P )y(t)\\
&=\sum_{i=1}^{N}y_{i}^{T}(t)\lambda_{i}Py_{i}(t)\\
&\geq\lambda_{2}\lambda_{\min}(P)\sum_{i=1}^{N}y_{i}^{T}(t)y_{i}(t)\\
&=\lambda_{2}\lambda_{\min}(P)\sum_{i=1}^{N}e_{i}^{T}(t)e_{i}(t)
\end{align*}

Additionally, by the condition $\lambda>\frac{1}{(1-\alpha)\rho(A)}$, we have
\begin{align*}
\sum_{i=1}^{N}e_{i}^{T}(t)Pd_{i}\rho_{i}(t)e_{i}(t)
>\sum_{i=1}^{N}e_{i}^{T}(t)Pd_{i}(\rho_{i}^{*}-\varepsilon)e_{i}(t).
\end{align*}

Therefore, one can obtain
\begin{align*}
\dot{V}_{1}(t)\leq e^{T}(t)\Big[(-\eta-\lambda_{2}\lambda_{\min}(P))I_{Nn}+I_{N}\otimes(P\Delta)-D\otimes P \Big]e(t),
\end{align*}
where
\begin{align*}
D=diag\{(d_{1}(\rho_{1}^{*}-\varepsilon),\cdots,d_{r_{l}}(\rho_{r_{l}}^{*}-\varepsilon),0,\cdots,0)\}
\end{align*}

Now, as for $V_{2}(t),$ we obtain the $V_{2i}(t)$'s differential along
the trajectories of Eq.~\eqref{infemodel}
\begin{align*}
\dot{V}_{2i}(t)&=\dot{S}_{i}(t)-\frac{S_{i}^{*}}{S_{i}(t)}\dot{S}_{i}(t)+\dot{\rho}_{i}(t)-\frac{\rho_{i}^{*}}{\rho_{i}(t)}\dot{\rho}_{i}(t)\\
&=\Big(1-\frac{S_{i}^{*}}{S_{i}(t)}\Big)\Big[\rho_{i}(t)-\lambda\varphi_{i}(t)(1-\rho_{i}(t))\sum_{j=1}^{N}a_{ij}\rho_{j}(t)\Big]\\
&\hspace*{1em}+\Big(1-\frac{\rho_{i}^{*}}{\rho_{i}(t)} \Big)\Big[ -\rho_{i}(t)+\lambda\varphi_{i}(t)(1-\rho_{i}(t))\sum_{j=1}^{N}a_{ij}\rho_{j}(t) \Big].
\end{align*}

Since $S_{i}^{*},\rho_{i}^{*},i=1,2,\cdots,N$, are the positive equilibrium of
the epidemic dynamics system~\eqref{infemodel}, e.g., there are following relations:
\begin{align*}
-\rho_{i}^{*}+\lambda\varphi_{i}(t)S_{i}^{*}\sum_{j=1}^{N}a_{ij}\rho_{j}^{*}=0
\end{align*}
and
\begin{align*}
S_{i}^{*}+\lambda\varphi_{i}(t)\sum_{j=1}^{N}a_{ij}\rho_{j}^{*}=1.
\end{align*}
Then, we have
\begin{align*}
\dot{V}_{2i}(t)&=\Big( 1-\frac{S_{i}^{*}}{S_{i}(t)} \Big)\Big[ S_{i}^{*}+\lambda\varphi_{i}(t)S_{i}^{*}\sum_{j=1}^{N}a_{ij}\rho_{j}^{*}-S_{i}(t)-\lambda\varphi_{i}(t)S_{i}(t)\sum_{j=1}^{N}a_{ij}\rho_{j}(t) \Big]\\
&\hspace*{1em}+\Big( 1-\frac{\rho_{i}^{*}}{\rho_{i}(t)} \Big)\Big[ -\rho_{i}^{*}+\lambda\varphi_{i}(t)S_{i}^{*}\sum_{j=1}^{N}a_{ij}\rho_{j}^{*}-\rho_{i}(t)+\lambda\varphi_{i}(t)S_{i}(t)\sum_{j=1}^{N}a_{ij}\rho_{j}(t)\Big]\\
&=\frac{1}{S_{i}(t)}(S_{i}(t)-S_{i}^{*})\Big[ -(S_{i}(t)-S_{i}^{*})+\lambda\varphi_{i}(t)S_{i}^{*}\sum_{j=1}^{N}a_{ij}\rho_{j}^{*}-\lambda\varphi_{i}(t)S_{i}(t)\sum_{j=1}^{N}a_{ij}\rho_{j}(t) \Big]-\rho_{i}^{*}\\
&\hspace*{1em}+\lambda\varphi_{i}(t)S_{i}^{*}\sum_{j=1}^{N}a_{ij}\rho_{j}^{*}-\rho_{i}(t)+\lambda\varphi_{i}(t)S_{i}(t)\sum_{j=1}^{N}a_{ij}\rho_{j}(t)
+\frac{{\rho_{i}^{*}}^{2}}{\rho_{i}(t)}-\lambda\varphi_{i}(t)\frac{\rho_{i}^{*}S_{i}^{*}}{\rho_{i}(t)}\sum_{j=1}^{N}a_{ij}\rho_{j}^{*}\\
&\hspace*{1em}+\rho_{i}^{*}
-\lambda\varphi_{i}(t)\frac{\rho_{i}^{*}S_{i}(t)}{\rho_{i}(t)}\sum_{j=1}^{N}a_{ij}\rho_{j}(t)\\
&=-\frac{1}{S_{i}(t)}\Big(S_{i}(t)-S_{i}^{*}\Big)^{2}+\lambda\sum_{j=1}^{N}a_{ij}\varphi_{i}(t)S_{i}^{*}\rho_{j}^{*}\Big[ 2-\frac{S_{i}^{*}}{S_{i}(t)}+\frac{\rho_{j}(t)}{\rho_{j}^{*}}-\frac{\rho_{i}(t)}{\lambda\sum_{j=1}^{N}a_{ij}\varphi_{i}(t)S_{i}^{*}\rho_{j}^{*}}\\
&\hspace*{1em}+\frac{{\rho_{i}^{*}}^{2}}{\lambda\sum_{j=1}^{N}a_{ij}\varphi_{i}(t)S_{i}^{*}\rho_{j}^{*}\rho_{i}(t)}-\frac{\rho_{i}^{*}}{\rho_{i}(t)}
-\frac{\rho_{i}^{*}S_{i}(t)\rho_{j}(t)}{\rho_{i}(t)S_{i}^{*}\rho_{j}^{*}}\Big]\\
&=-\frac{1}{S_{i}(t)}\Big(S_{i}(t)-S_{i}^{*}\Big)^{2}+\lambda\sum_{j=1}^{N}a_{ij}\varphi_{i}(t)S_{i}^{*}\rho_{j}^{*}\Big[ 2-\frac{S_{i}^{*}}{S_{i}(t)}+\frac{\rho_{j}(t)}{\rho_{j}^{*}}-\frac{\rho_{i}(t)}{\rho_{i}^{*}}-\frac{\rho_{i}^{*}S_{i}(t)\rho_{j}(t)}{\rho_{i}(t)S_{i}^{*}\rho_{j}^{*}}\Big]
\end{align*}
Set $\tilde{a}_{ij}=a_{ij}\varphi_{i}(t)S_{i}^{*}\rho_{j}^{*},$ and
\begin{align*}
F_{ij}(S_{i}(\cdot),\rho_{i}(\cdot),\rho_{j}(\cdot))=2-\frac{S_{i}^{*}}{S_{i}(t)}+\frac{\rho_{j}(t)}{\rho_{j}^{*}}-\frac{\rho_{i}(t)}{\rho_{i}^{*}}-\frac{\rho_{i}^{*}S_{i}(t)\rho_{j}(t)}{\rho_{i}(t)S_{i}^{*}\rho_{j}^{*}}.
\end{align*}
Let
\begin{align*}
G_{i}(\rho_{i})=-\frac{\rho_{i}(t)}{\rho_{i}^{*}}+\ln\frac{\rho_{i}(t)}{\rho_{i}^{*}}
\end{align*}
\begin{align*}
\Phi(a)=1-a+\ln a
\end{align*}
It is obvious that the function $\Phi(a)$ is non-positive, i.e., $\Phi(a)\leq0,$ and $\Phi(a)=0$ if and only if $a=1.$
Then we can get that
\begin{align*}
F_{ij}(S_{i}(\cdot),\rho_{i}(\cdot),\rho_{j}(\cdot))&=G_{i}(\rho_{i})-G_{j}(\rho_{j})+\Phi\Big(\frac{S_{i}^{*}}{S_{i}(t)}\Big)+\Phi\Big(\frac{\rho_{i}^{*}S_{i}(t)\rho_{j}(t)}{\rho_{i}(t)S_{i}^{*}\rho_{j}^{*}}\Big)\\
&\leq G_{i}(\rho_{i})-G_{j}(\rho_{j})
\end{align*}

According to the theorem in reference~\cite{paper37}, we can claim that $V_{2}(t)=\sum_{i=1}^{N}c_{i}V_{2i}(t)$ is a \emph{Lyapunov} function, for which $\dot{V}_{2}(t)\leq0$, and $\dot{V}_{2}(t)=0$ if and only if $S_{i}(t)=S_{i}^{*},\rho_{i}(t)=\rho_{i}^{*},i=1,2,\cdots,N.$ Which means that the invariant set of $\dot{V}_{2}(t)=0$ is a single point set $\{ E^{*} \}=\{(S_{1}^{*},\rho_{1}^{*},\cdots,S_{N}^{*},\rho_{N}^{*})\}.$

To sum up, when the condition~\eqref{condition1} hold, the \emph{Lyapunov} function $\dot{V}(t)<0$. Which indicates that, when infectious disease spreads, i.e., $\lambda>\frac{1}{(1-\alpha)\rho(A)},$  the corresponding endemic equilibrium $E^{*}$ is globally asymptotically stable. Meanwhile, under the controllers~\eqref{controller1}, the behavioral dynamics system~\eqref{infenetmodel} achieves behavior synchronization.

\end{proof}
\end{theorem}

\subsection*{B. Control the infected and susceptible individuals}

In this part, infected and uninfected individuals are both under control,
which eventually leads individuals in each group based on distinct information about the individuals'
diseases to reach diverse behaviors.

Dividing all individuals into two groups,
infected and uninfected, as shown in the following:
$$\{\underbrace{1,~2,~\cdots,~r_{1},}_{\text{infected individuals}}\underbrace{r_{1}+1,r_{1}+2,\cdots,r_{2}}_{\text{Uninfected individuals}}\}.$$
Or represented in the set form as
\begin{align*}
\mathbb{K}=\{r_{k-1}+1,r_{k-1}+2,\cdots,r_{k}\},
\end{align*}
in which $k=1,2,$ and $r_{0}=0,r_{2}=N.$

Blocking topology structure of the network system, we obtain
\begin{align*}
L&=\left[
\begin{array}{cccc}
  l_{11} & l_{12} & \cdots & l_{1N} \\
  l_{21} & l_{22} & \cdots & l_{1N} \\
  \vdots & \vdots & \ddots & \vdots \\
  l_{N1} & l_{N2} & \cdots & l_{NN}
\end{array}
\right]\\
\\
&=\left[
    \begin{array}{cc}
      L_{11} ~& ~L_{12} \\ \\
      L_{12} ~& ~L_{22} \\
    \end{array}
  \right]
\end{align*}
where $L_{kk}\in{\mathbb{R}^{(r_{k}-r_{k-1})\times(r_{k}-r_{k-1})}}$ belongs to $A2,$ $L_{kq}\in\mathbb{R}^{(r_{k}-r_{k-1})\times q_{k}},q_{k}=N-(r_{k}-r_{k-1}),$ belongs to $A3.$

Then, we design the controllers for infected individuals and uninfected
individuals, respectively, as follows:
\begin{align}\label{controller2}
u_{i}(t)=\left\{
\begin{aligned}
&-d_{i}\rho_{i}(t)(x_{i}(t)-s_{1}(t)),\hspace*{3.7em} i\in\{1,2,\cdots,r_{1}\},\\
&-\beta\xi_{i}S_{i}(t)(x_{i}(t)-s_{2}(t)),\hspace*{3em} i\in\{r_{1}+1,r_{1}+2,\cdots,r_{2}\}.
\end{aligned}
\right.
\end{align}
where $d_{i}$ is the infected controlling strength added to $i$-th infected individual,
and $d_{i}>0$ when $i=1,2,\cdots,r_{s},$ otherwise $d_{i}=0.$ $\xi_{i}$ is the uninfected
controlling strength to susceptible individuals, and $\xi_{i}>0$
when $i=r_{1}+1,r_{1}+2,\cdots,r_{1}+s,$ otherwise $\xi_{i}=0.$ $\beta$ is the global
coefficient of elasticity for susceptible individuals.

\begin{remark}
In this control strategy, individuals in the complex network system will reach
two desired consistent behaviors. On the one hand, individuals in the infected group will
go to hospital to receive treatment or to be isolated from getting in touch with healthy
individuals. On the other hand, the susceptible individuals in the uninfected group are going to wash their
hands frequently or have a rest to keep nice health state. What's more, they are informed
to avoid staying at crowded place for long time and to open the window maintaining good indoor
ventilation. Namely, the epidemic behavioral cluster synchronization is reached for two groups based
on the information about individuals' infection.
\end{remark}

Therefore, we consider the synchronization state as follows:
\begin{align}\label{target2}
\dot{s}_{k}(t)=f(s_{k}(t)).
\end{align}
Then the corresponding errors can be designed as
$$e_{i}(t)=x_{i}(t)-s_{k}(t),\quad i\in\mathbb{K},\quad k=1,2.$$

Before theoretical analysis, we give some notations that will be used later.
Let $\mathbb{Q}=\{1,2,\cdots,r_{k-1},r_{k}+1,\cdots,N\}.$ $\widetilde{e}_{j}^{(k)}(t)=(e_{r_{k-1}+1,j}(t),e_{r_{k-1}+2,j}(t),\cdots,e_{r_{k},j}(t))^{T}\in\mathbb{R}^{r_{k}-r_{k-1}},$ and $\widetilde{e}_{j}^{(q)}(t)=(e_{1,j}(t),\cdots,e_{r_{k-1}j}(t),\\e_{r_{k}+1,j}(t),\cdots,e_{N,j}(t))^{T}\in\mathbb{R}^{N-(r_{k}-r_{k-1})}.$

\begin{theorem}\label{synctheorem2}
Assume that the local dynamics $f$ belongs to the QUAD function class. When the infection
rate $\lambda>\frac{1}{(1-\alpha)\rho(A)},$ which indicates that the disease
may spread, then the endemic equilibrium $E^{*}$ is globally asymptotically stable, as well as
the behavioral dynamics system~\eqref{infenetmodel} will achieve behavioral
cluster synchronization for two groups based on the information about individuals'
infection under the grouped control strategies~\eqref{controller2} if there exists a
positive $\theta>0$ such that:
\begin{align}\label{condition2}
\Big(\frac{\zeta}{\lambda_{\min}(P)}+\frac{\theta}{2}-d_{k}^{*}\Big)I_{r_{k}-r_{k-1}}-L_{kk}+\frac{1}{2\theta}L_{kq_{k}}L_{kq_{k}}^{T}<0, \quad k=1,2,
\end{align}
where $d_{1}^{*}=\min\limits_{i\in\{1,2,\cdots,r_{s} \}}\{d_{i}(\rho_{i}^{*}-\varepsilon)\},$ and $d_{2}^{*}=\min\limits_{i\in\{r_{1}+1,\cdots,r_{1}+s \}}\{\beta\xi_{i}(1-\rho_{i}^{*}-\varepsilon)\}.$

\begin{proof}
Define the Lyapunov function as follows,
\begin{align}
V(t)=V_{2}(t)+V_{3}(t),
\end{align}
where $V_{2}(t)$ has the form shown as
\begin{align*}
V_{2}(t)=\sum_{k=1}^{2}V_{2}^{(k)}(t),
\end{align*}
in which
\begin{align*}
V_{2}^{(k)}(t)=\sum_{i\in \mathbb{K}}c_{i}^{(k)}V_{2i}^{(k)},
\end{align*}
and consisting with the case of different groups, $V_{2i}^{(k)}$ is defined as
\begin{align*}
V_{2i}^{(k)}=S_{k,i}(t)-S_{k,i}^{*}-S_{k,i}^{*}\ln\frac{S_{k,i}(t)}{S_{k,i}^{*}}+\rho_{k,i}(t)-\rho_{k,i}^{*}\ln\frac{\rho_{k,i}(t)}{\rho_{k,i}^{*}}.
\end{align*}
And also $S_{k,i}(t),$ $\rho_{k,i}(t)$ are supposed to satisfy the corresponding epidemic
models as
\begin{align*}
\left\{
\begin{aligned}
&\dot{\rho}_{k,i}(t)=-\rho_{k,i}(t)+\lambda\varphi_{k,i}(t)(1-\rho_{k,i}(t))\sum_{j=1}^{N}a_{ij}\rho_{j}(t)\\
&\dot{S}_{k,i}(t)=\rho_{k,i}(t)-\lambda\varphi_{k,i}(t)S_{k,i}(t)\sum_{j=1}^{N}a_{ij}\rho_{j}(t).
\end{aligned}
\right.
\end{align*}
And the analysis for $V_{2i}^{(k)}(t)$ is similar as in Theorem~\ref{synctheorem} for $V_{2i}(t),$ and have the same results that $V_{2}^{(k)}(t)$ is a \emph{Lyapunov} function, for which $\dot{V}_{2}^{(k)}(t)\leq0$ and $\dot{V}_{2}^{(k)}(t)=0$ if and only if $S_{k,i}(t)=S_{k,i}^{*},$ $\rho_{k,i}(t)=\rho_{k,i}^{*},$ $i\in\mathbb{K},k=1,2.$ Then, $V_{2}(t)$ can be claimed to be a \emph{Lyapunov} function, and the invariant set of $\dot{V}_{2}(t)=0$ is the single point set written as $\{(S_{1,1}^{*},\rho_{1,1}^{*},\cdots,S_{1,r_{1}}^{*},\rho_{1,r_{1}}^{*},S_{2,r_{1}+1}^{*},\rho_{2,r_{1}+1}^{*},\cdots,S_{2,r_{2}}^{*},\rho_{2,r_{2}}^{*})\}.$

Then, $V_{3}(t)$ is also selected as~\eqref{V1}, but has another equivalence form for cluster analysis,
\begin{align}
V_{3}(t)=\frac{1}{2}\sum_{k=1}^{2}\sum_{i\in\mathbb{K}}e_{i}^{T}(t)Pe_{i}(t).
\end{align}

At first, differentiate the $V_{3}(t)$
\begin{align*}
\dot{V}_{3}(t)&=\sum_{k=1}^{2}\sum_{i\in\mathbb{K}}e_{i}^{T}(t)P\dot{e}_{i}(t)\\
&=\sum_{k=1}^{2}\sum_{i\in\mathbb{K}}e_{i}^{T}(t)P[f(x_{i}(t))-f(s_{k}(t))-\sum_{j=1}^{N}l_{ij}e_{j}(t)+u_{i}(t)]\\
&=\sum_{k=1}^{2}\sum_{i\in\mathbb{K}}e_{i}^{T}(t)P[f(x_{i}(t))-f(s_{k}(t))]-\sum_{i=1}^{N}\sum_{j=1}^{N}e_{i}^{T}(t)Pl_{ij}e_{j}(t)+\sum_{i=1}^{T}e_{i}^{T}(t)Pu_{i}(t)
\end{align*}

Next, we analyze each term in the aspect of clustering shown above.

For the first term, from the property of QUAD function class, we have
\begin{align}\label{firstterm}
&\sum_{k=1}^{2}\sum_{i\in\mathbb{K}}e_{i}^{T}(t)P[f(x_{i}(t))-f(s_{k}(t))]\\
&\leq\zeta\sum_{k=1}^{2}\sum_{i\in\mathbb{K}}e_{i}^{T}(t)e_{i}(t)\notag\\
&\leq\frac{\zeta}{\lambda_{\min}(P)}\sum_{k=1}^{2}\sum_{i\in\mathbb{K}}e_{i}^{T}(t)Pe_{i}(t)\notag\\
&=\frac{\zeta}{\lambda_{\min}(P)}\sum_{k=1}^{2}\sum_{i\in\mathbb{K}}\sum_{j=1}^{n}e_{ij}(t)p_{j}e_{ij}\notag\\
&=\frac{\zeta}{\lambda_{\min}(P)}\sum_{k=1}^{2}\sum_{j=1}^{n}({\widetilde{e}_{j}^{(k)}(t)})^{T}\widetilde{e}_{j}^{(k)}(t).
\end{align}

As for the second term, it may be obtained that
\begin{align*}
-\sum_{i=1}^{N}\sum_{j=1}^{N}e_{i}^{T}(t)Pl_{ij}e_{j}(t)
&=-\sum_{k=1}^{2}\Big[\sum_{i,j\in\mathbb{K}}e_{i}^{T}(t)Pl_{ij}e_{j}(t)+\sum_{i\in\mathbb{K},j\in\mathbb{Q}}e_{i}^{T}(t)Pl_{ij}e_{j}(t)\Big]\\
&=-\sum_{k=1}^{2}\Big[ \sum_{i,j\in\mathbb{K}}l_{ij}\sum_{l=1}^{n}p_{l}e_{il}(t)e_{jl}(t)+\sum_{i\in\mathbb{K},j\in\mathbb{Q}}l_{ij}\sum_{l=1}^{n}e_{il}(t)p_{l}e_{jl}(t) \Big]\\
&=-\sum_{k=1}^{2}\Big[ \sum_{l=1}^{n}p_{l}\sum_{i,j\in\mathbb{K}}e_{il}(t)l_{ij}e_{jl}(t)+\sum_{l=1}^{n}p_{l}\sum_{i\in\mathbb{K},j\in\mathbb{Q}}e_{il}(t)l_{ij}e_{jl}(t)\Big]\\
&=-\sum_{k=1}^{2}\Big[ \sum_{l=1}^{n}p_{l}({\widetilde{e}_{l}^{(k)}(t)})^{T}L_{kk}\widetilde{e}_{l}^{(k)}(t)+\sum_{l=1}^{n}p_{l}(\widetilde{e}_{l}^{(k)}(t))^{T}L_{kq_{k}}\widetilde{e}_{l}^{(q)}(t)\Big],
\end{align*}
according to the Lemma~\ref{L1} $(x^{T}y\leq\frac{1}{2\theta}x^{T}x+\frac{\theta}{2}y^{T}y, \text{existing~} \theta>0)$, we obtain the following inequality
\begin{align*}
-\sum_{i=1}^{N}\sum_{j=1}^{N}e_{i}^{T}(t)Pl_{ij}e_{j}(t)\leq-\sum_{k=1}^{2}\sum_{j=1}^{n}p_{j}(\widetilde{e}_{j}^{(k)}(t))^{T}L_{kk}\widetilde{e}_{j}^{(k)}(t)\\
+\sum_{k=1}^{2}\sum_{j=1}^{n}p_{j}\Big[\frac{1}{2\theta}(\widetilde{e}_{j}^{(k)}(t))^{T}L_{kq_{k}}L_{kq_{k}}^{T}\widetilde{e}_{j}^{(k)}(t)+\frac{\theta}{2}(\widetilde{e}_{j}^{(q)}(t))^{T}\widetilde{e}_{j}^{(q)}(t)\Big]
\end{align*}

Note that $(\widetilde{e}_{j}^{(q)}(t))^{T}\widetilde{e}_{j}^{(q)}(t)=\sum_{k=1}^{2}(\widetilde{e}_{j}^{(k)}(t))^{T}\widetilde{e}_{j}^{(k)}(t)-(\widetilde{e}_{j}^{(k)}(t))^{T}\widetilde{e}_{j}^{(k)}(t),$ then we get that
\begin{align}\label{secondterm}
-\sum_{i=1}^{N}\sum_{j=1}^{N}e_{i}^{T}(t)Pl_{ij}e_{j}(t)
&\leq-\sum_{k=1}^{2}\sum_{j=1}^{n}p_{j}(\widetilde{e}_{j}^{(k)}(t))^{T}L_{kk}\widetilde{e}_{j}^{(k)}(t)\notag\\
&\hspace*{1em}+\frac{1}{2\theta}\sum_{k=1}^{2}\sum_{j=1}^{n}p_{j}(\widetilde{e}_{j}^{(k)}(t))^{T}L_{kq_{k}}L_{kq_{k}}^{T}\widetilde{e}_{j}^{(k)}(t)
+\frac{\theta}{2}\sum_{k=1}^{2}\sum_{j=1}^{n}p_{j}(\widetilde{e}_{j}^{(k)}(t))^{T}\widetilde{e}_{j}^{(k)}(t).
\end{align}

Considering the third term, we can get
\begin{align*}
\sum_{i=1}^{N}e_{i}^{T}(t)Pu_{i}(t)=-\sum_{i=1}^{r_{1}}e_{i}^{T}(t)P d_{i}\rho_{i}(t)(x_{i}(t)-s_{1}(t))-\sum_{i=r_{1}+1}^{N}e_{i}^{T}(t)P\beta\xi_{i}(1-\rho_{i}(t))(x_{i}(t)-s_{2}(t))
\end{align*}

From the fact that when $\lambda>\frac{1}{(1-\alpha)\rho(A)},$ there is a unique endemic equilibrium $E^{*}$ for the epidemic dynamics system~\eqref{infemodel}, which means that there exists an $\varepsilon>0,$ and $t_{0},$ when $t\geq t_{0},$ then $|\rho_{i}(t)-\rho_{i}^{*}|<\varepsilon.$
Therefore,
\begin{align}\label{thirdterm}
\sum_{i=1}^{N}e_{i}^{T}(t)Pu_{i}(t)&=-\sum_{i=1}^{r_{1}}\sum_{j=1}^{n}d_{i}\rho_{i}(t)e_{ij}(t)p_{j}e_{ij}(t)-\sum_{r_{1}+1}^{N}\sum_{j=1}^{n}\beta\xi_{i}(1-\rho_{i}(t))e_{ij}(t)p_{j}e_{ij}(t)\notag\\
&\leq-\sum_{i=1}^{r_{1}}\sum_{j=1}^{n}d_{i}(\rho_{i}^{*}-\varepsilon)e_{ij}(t)p_{j}e_{ij}(t)-\sum_{i=r_{1}+1}^{N}\sum_{j=1}^{n}\beta\xi_{i}(1-\rho_{i}^{*}-\varepsilon)e_{ij}(t)p_{j}e_{ij}(t)\notag\\
&\leq-\sum_{j=1}^{n}p_{j}\sum_{i=1}^{r_{1}}e_{ij}(t)d_{1}^{*}e_{ij}(t)-\sum_{j=1}^{n}p_{j}\sum_{i=r_{1}+1}^{N}e_{ij}(t)d_{2}^{*}e_{ij}(t)\notag\\
&=-d_{1}^{*}\sum_{j=1}^{n}(\widetilde{e}_{j}^{(1)}(t))^{T}\widetilde{e}_{j}^{(1)}(t)-d_{2}^{*}\sum_{j=1}^{n}(\widetilde{e}_{j}^{(2)}(t))^{T}\widetilde{e}_{j}^{(2)}(t)\notag\\
&=-\sum_{k=1}^{2}\sum_{j=1}^{n}p_{j}d_{k}^{*}(\widetilde{e}_{j}^{(k)}(t))^{T}\widetilde{e}_{j}^{(k)}(t)
\end{align}
where $d_{1}^{*}=\min\limits_{i\in\{1,2,\cdots,r_{s}\}}\{d_{i}(\rho_{i}^{*}-\varepsilon)\},$ $d_{2}^{*}=\min\limits_{i\in\{r_{1}+1,\cdots,r_{1}+s\}}\{\beta\xi_{i}(1-\rho_{i}^{*}-\varepsilon)\}.$

Hereto, combining with~\eqref{firstterm}, \eqref{secondterm} and~\eqref{thirdterm}, we obtain
\begin{align}
\dot{V}_{3}(t)\leq\sum_{k=1}^{2}\sum_{j=1}^{n}p_{j}(\widetilde{e}_{j}^{(k)}(t))^{T}\Big[\Big(\frac{\zeta}{\lambda_{\min}(P)}+\frac{\theta}{2}-d_{k}^{*}\Big)I_{r_{k}-r_{k-1}}-L_{kk}+\frac{1}{2\theta}L_{kq_{k}}L_{kq_{k}}^{T}\Big]\widetilde{e}_{j}^{(k)}(t)
\end{align}


In conclusion, we get
\begin{align*}
\dot{V}(t)&=\dot{V}_{2}(t)+\dot{V}_{3}(t)<0
\end{align*}
if conditions~\eqref{condition2} are satisfied.

Therefore, we can claim that when $\lambda>\frac{1}{(1-\alpha)\rho(A)},$ resulting in the
infectious disease transmission, then the endemic equilibrium is globally asymptotically
stable. And under the control strategy~\eqref{controller2}, only when
conditions~\eqref{condition2} are satisfied, the behavioral
dynamics system~\eqref{infenetmodel} reaches behavioral
cluster synchronization for two groups
based on the disease information. Then the proof is completed.

\end{proof}
\end{theorem}

\section{Numerical Simulation}\label{simulation}

In this section, we shall give two numerical simulation examples to demonstrate the obtained
main results in this paper.
At first, using the NW small-world principle with the average degree as $4$ and the
rewiring probability as $0.1$, and the average degree as $6$ and the rewiring probability
also as $0.1$, to generate two parts with size of $20$ nodes and $30$ nodes, respectively.
And those two parts are combined by the rules that existing $2$ edges from the first part
to the second part and $3$ edges from the second to the first one. Which generates the
considered network topology with the node's dimension as $2$.

\begin{exa}\label{exa1}
Then, as for the first control strategy, we select the local dynamics
as $$f(x)=C\cdot x+A\cdot tanhx,$$ where each node has dimension $2$ and
\begin{align*}
C=\left(
    \begin{array}{cc}
      -1 & 0 \\
      0 & -1 \\
    \end{array}
  \right),\qquad
A_1=\left(
    \begin{array}{cc}
      -1 & -0.5 \\
      -5 & 4.5 \\
    \end{array}
  \right).
\end{align*}
And we select the matrices $P$ and $\Delta$ as
\begin{align*}
P=\left(
    \begin{array}{cc}
      1 & 0 \\
      0 & 1 \\
    \end{array}
  \right),\qquad
\Delta=0.5\times\left(
         \begin{array}{cc}
           1 & 0 \\
           0 & 1 \\
         \end{array}
       \right).
\end{align*}
Then we can select $\eta=1.4394$ such that
\begin{align*}
(x-y)^{T}P[f(x)-f(y)-\Delta(x-y)]\leq-\eta(x-y)^{T}(x-y).
\end{align*}
According to the considered network, we can get the $\lambda_2=0.235.$
Therefore, under the synchronous condition in Theorem~\ref{synctheorem},
by simple computing, we can get that
\begin{align*}
-\eta-\lambda_{2}\lambda_{\min}(P)+0.5<0
\end{align*}
So, it is only needed to select the appropriate control strength and the synchronous
condition is also satisfied naturally.
From the network's topology, we can compute the $\rho(A)=8.2574.$ Let $\alpha=0.5,$
then the epidemic threshold can be
obtained as $\lambda_c=\frac{1}{(1-\alpha)\rho(A)}=0.2422,$
while $\frac{1}{(1+\alpha)\rho(A)}=0.0807.$

Next, let the initial epidemic probability $\rho_0=0.01,$ and select the appropriate
initial values for nodes. And adding the controller to the first $10$ nodes with the
strength $d_{i}=20,i=1,\cdots,10.$ We set the infection rate $\lambda=0.3>\lambda_c.$
Then, we can obtain the Fig.~\ref{0.3-rho-E}.
\begin{figure}[htbp]
  \centering
  \includegraphics[width=3.5in]{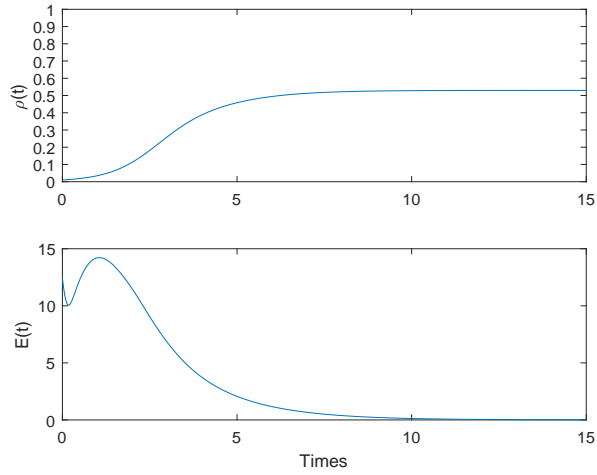}
  \caption{The epidemic probability and the synchronization errors for all nodes under the first control strategy with the infection rate $\lambda=0.3.$}\label{0.3-rho-E}
\end{figure}
From which one can see that when the infection rate is larger than the
epidemic threshold, there exists a positive stable value in the end referred by the average
infected density $\rho(t)$'s ($\rho(t)=\sum_{i=1}^{50}\rho_{i}(t)$) curve, which means the
disease will spread and will continuously exist to be the endemic disease. In this case,
under the feedback control strategy, the error picture tends to zero which reflects that
the network achieve behavior synchronization arouse by epidemic transmission. If we
select $\lambda=0.7,$ then there will be a larger stable value for $\rho(t)$
as shown in Fig.~\ref{0.7-rho-E}, and the behavior synchronization resulting from
disease transmission also emerge under the pinning feedback controllers.
\begin{figure}[htbp]
  \centering
  \includegraphics[width=3.5in]{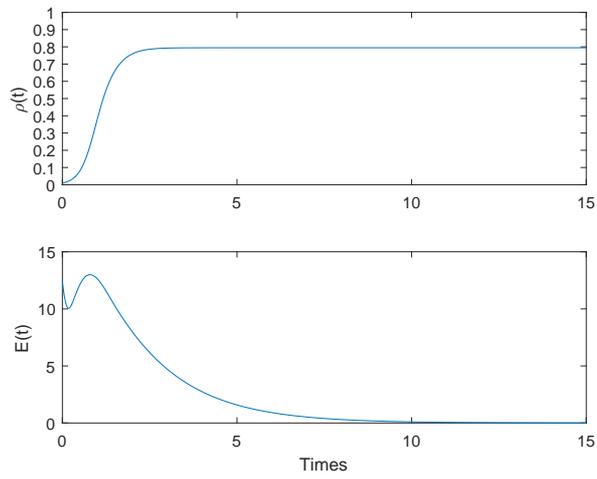}
  \caption{The epidemic probability and the synchronization errors for all nodes under the first control strategy with the infection rate $\lambda=0.7.$}\label{0.7-rho-E}
\end{figure}

When select the infection rate less than the value $\frac{1}{(1+\alpha)\rho(A)}$,
such as $\lambda=0.06<0.0807.$
\begin{figure}[htbp]
  \centering
  \includegraphics[width=3.5in]{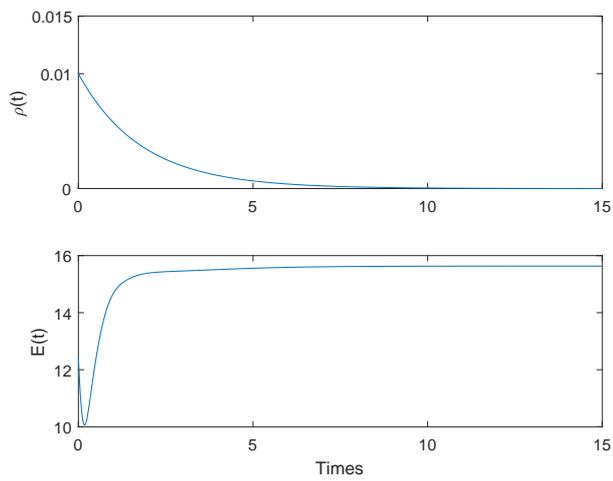}
  \caption{The epidemic probability and the synchronization errors for all nodes under the first control strategy with the infection rate $\lambda=0.06.$}\label{0.06-rho-E}
\end{figure}
From the Fig.~\ref{0.06-rho-E} one can see that the $\rho(t)$ reach to zero in the end,
which means the disease shall be extinct eventually. And the network will not achieve
behavior synchronization with the error going to none zero result from the invalid controllers,
as shown in the error curve.
\end{exa}

\begin{exa}\label{exa2}
Now, we assume that the first $20$ nodes constitute the infectious individual group in the
above concerned network, and the latter $30$ ones represent the susceptible group, we shall
show that when the epidemic spreads, the network can reach two different behavior
synchronization states for those two groups individuals via the second control strategy
in this paper. In order to demonstrate this case, we introduce two types of local dynamical
behaviors with the same form as Example~\ref{exa1}, but distinct coefficient matrices
as
\begin{align*}
C_2=\left(
     \begin{array}{cc}
        -4.5 & 0 \\
         0   & -4.5
     \end{array}
     \right),
A_2=\left(
     \begin{array}{cc}
       -2 & -0.1 \\
       -5 & 4.5 \\
     \end{array}
   \right)~
\text{and}~
C_3=\left(
     \begin{array}{cc}
       -1.8 & 0 \\
        0   & -1.8
     \end{array}
     \right),
A_3=\left(
     \begin{array}{cc}
       0.1 & -0.1 \\
       -0.4 & 0.4 \\
     \end{array}
   \right).
\end{align*}
Then, we can get the $\zeta=2.4462$ such that $$(x-y)^{T}[f(x)-f(y)]\leq\zeta(x-y)^{T}(x-y).$$
Select $\theta=1,$ and we can easily get that
\begin{align*}
\lambda_{\max}\big(-L_{11}+\frac{1}{2}L_{12}L_{12}^{T}\big)<-\big(\zeta+\frac{1}{2}\big),\\
\lambda_{\max}\big(-L_{22}+\frac{1}{2}L_{21}L_{21}^{T}\big)<-\big(\zeta+\frac{1}{2}\big),
\end{align*}
so the synchronization conditions in Theorem~\ref{synctheorem2} are satisfied under the
appropriate control strength. In this case, we select the initial infectious probabilities
for two groups as $\rho_{1}(0)=0.01$ and $\rho_{2}(0)=0.04,$ respectively.

Then let $\lambda=0.4>0.2422,$ by using the second control strategy, adding the
controllers to a part of nodes in each group with the appropriate control strength,
the epidemic probability reach a stable state, and the network achieve two types of behavior
synchronization. In Fig.~\ref{0.4-rho-E1E2-E-E12}, one can see that the $\rho_{1}(t)$ ($\rho_{1}(t)=\sum_{i=1}^{20}\rho_{i}(t)$) and $\rho_{2}(t)$ ($\rho_{2}(t)=\sum_{i=1}^{30}\rho_{i}(t)$) get the positive stable values
in the end. And the errors $E1(t)$ and $E2(t)$ in each groups and the whole errors $E(t)$
of all nodes' states are also tend to the zeros, while the errors $E12(t)$ between two
groups do not tend to zero, which indicates the main results in our paper.
\begin{figure}[htbp]
  \centering
  \includegraphics[width=4in]{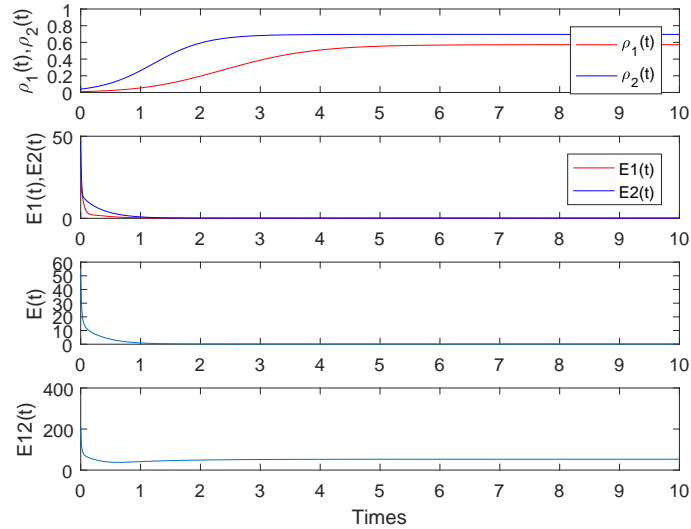}
  \caption{The $\rho_{1}(t)$, $\rho_{2}(t)$, the errors in each groups, all errors of the whole nodes' states in the network and the errors between each groups with the infection rate $\lambda=0.4$ under the second control strategy. }\label{0.4-rho-E1E2-E-E12}
\end{figure}
If we select $\lambda=0.7,$ then the infectious probability for each groups will have
a higher stable values, as shown in Fig.~\ref{0.7-rho-E1E2-E-E12}.
\begin{figure}[htbp]
  \centering
  \includegraphics[width=4in]{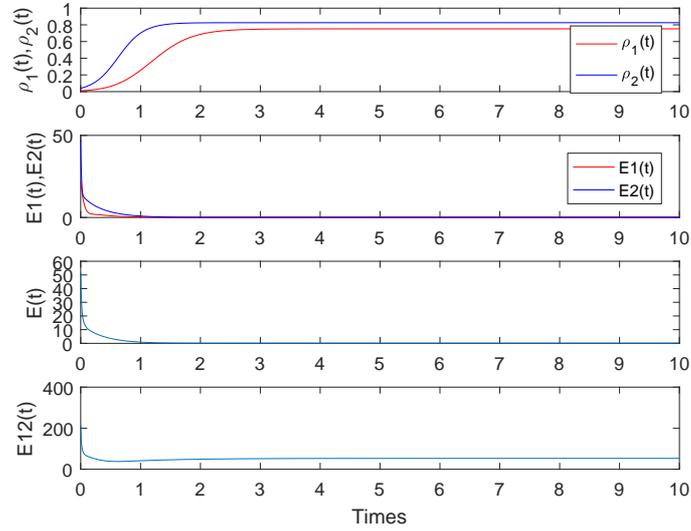}
  \caption{The $\rho_{1}(t)$, $\rho_{2}(t)$, the errors in each groups, all errors of the whole nodes' states in the network and the errors between each groups with the infection rate $\lambda=0.7$ under the second control strategy. }\label{0.7-rho-E1E2-E-E12}
\end{figure}

When the infection rate is less than the value $\frac{1}{(1+\alpha)\rho(A)}$, e.g., $\lambda=0.07<0.0807,$ we can obtain Fig.~\ref{0.07-rho-E1E2-E-E12}. In which,
the infectious probabilities for two groups $\rho_{1}(t)$ and $\rho_{2}(t)$ tend to
zero eventually. And the errors in two groups $E1(t)$ and $E2(t),$ as well as the
whole errors for the network do not tend to zero. Which indicates that once the
infection rate is less than a certain value $\frac{1}{(1+\alpha)\rho(A)}$,
the disease will be extinct in the end, and meanwhile the controllers are invalid so
that none behavior synchronization.
\begin{figure}[htbp]
  \centering
  \includegraphics[width=4in]{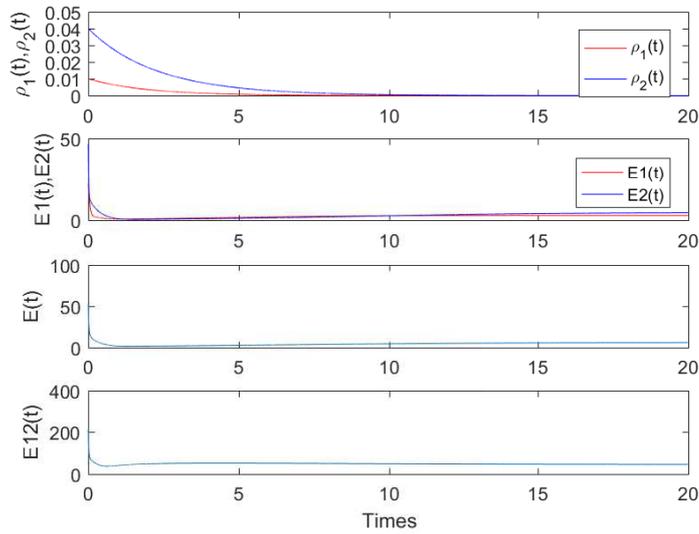}
  \caption{The $\rho_{1}(t)$, $\rho_{2}(t)$, the errors in each groups, all errors of the whole nodes' states in the network and the errors between each groups with the infection rate $\lambda=0.07$ under the second control strategy. }\label{0.07-rho-E1E2-E-E12}
\end{figure}
\end{exa}

\begin{remark}
As for the case that the infection rate is located between the
value $\frac{1}{(1+\alpha)\rho(A)}$ and the epidemic threshold $\frac{1}{(1-\alpha)\rho(A)},$
we select two values of $\lambda$ to see what changes there will be under the two control strategies.
On the one hand, let $\lambda=0.15\in[0.0807,0.2422].$ Under the first control strategy,
the $\rho(t)$ curve and errors are shown in Fig.~\ref{0.15-rho-E}. We can see that
the infectious probability tends to a stable positive value, while the network dose not
get the behavior synchronization reflected by the error curve. On the other hand,
let $\lambda=0.1,$ and the infectious possibilities and the errors under the second
control strategy are shown in Fig.~\ref{0.1-rho-E1E2-E-E12}. In that case, the
infection rate tend to zero, and the errors so not. Hence, it may be
more complex and varied in the case that the infection rate belongs
to the interval $[0.0807,0.2422],$ and some corresponding measures should be discussed
according to the concrete conditions.
\begin{figure}[htbp]
  \centering
  \includegraphics[width=4in]{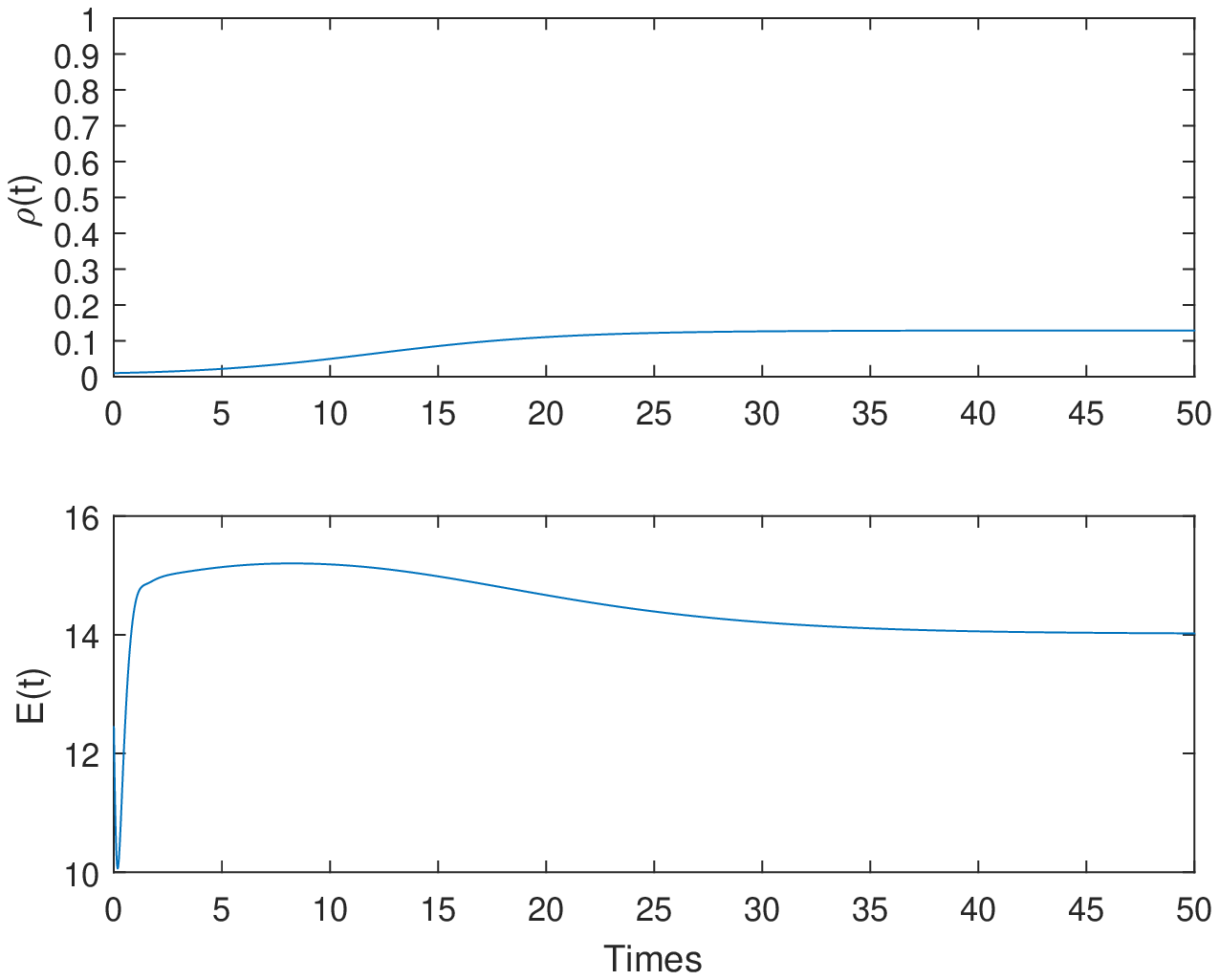}
  \caption{The epidemic probability and the synchronization errors for all nodes under the first control strategy with the infection rate $\lambda=0.15.$}\label{0.15-rho-E}
\end{figure}

\begin{figure}[htbp]
  \centering
  \includegraphics[width=4in]{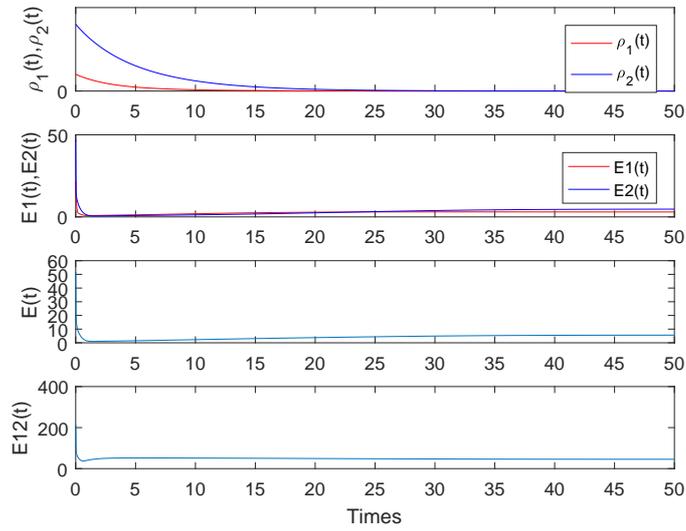}
  \caption{The $\rho_{1}(t)$, $\rho_{2}(t)$, the errors in each groups, all errors of the whole nodes' states in the network and the errors between each groups with the infection rate $\lambda=0.1$ under the second control strategy. }\label{0.1-rho-E1E2-E-E12}
\end{figure}
\end{remark}

\section{Conclusions}\label{conclusions}

In this paper, we investigate the epidemic controlling behavioral synchronization problem
for a class of epidemic synchronous complex dynamical networks aroused from the contagions
transmission, in which the quenched mean-field theory is introduced in the epidemic system.
More realistically, the case of inhibition of contact behavior is concerned in the epidemic
models. First, the existence of endemic equilibrium in view of the epidemic threshold and the
global stability of disease-free equilibrium are discussed. Then, based on the proven
technique of pinning feedback strategy, where only a part of nodes need to be controlled,
and the design of two types of pinning schemes depending on nodes' information of infection,
two desired epidemic control behavior synchronization are achieved by means of the
definition of behavioral synchronization. Sufficient conditions for epidemic control
behavior synchronization under the contagions spreading are derived. The numerical
simulations are exhibited to demonstrate the effectiveness of the main results obtained
in this paper.

\section*{Acknowledgments}
\indent
This work was jointly supported by the NSFC under grants 11572181 and 11331009.

\end{document}